\documentclass[useAMS,referee,usenatbib]{biom} 

\usepackage{amsmath}
\usepackage{amssymb}
\usepackage[dvips]{epsfig}
\usepackage{dcolumn}
\usepackage{enumerate}
\usepackage{hhline}
\usepackage{dsfont}
\usepackage{afterpage}
\usepackage{arydshln}
\usepackage{graphicx}
\usepackage{color}
\usepackage[usenames,dvipsnames]{xcolor}
\usepackage{rotating}
\usepackage[breaklinks]{hyperref}
\usepackage[]{caption}
\usepackage{placeins}
\usepackage{algorithmicx}
\usepackage[noend]{algpseudocode}
\usepackage{algorithm}
\usepackage{diagbox}
\usepackage{graphicx}
\usepackage{wrapfig}
\usepackage{lscape}
\input epsf
\usepackage{fontenc}
\usepackage{setspace}
\usepackage{bm}
\usepackage{slashbox}
\usepackage{lscape}
\usepackage{breakurl} 
\usepackage{multirow}
\usepackage{eurosym}
\usepackage{titlesec}

\epsfverbosetrue

\def \dsE {\text{$\mathds{E}$}}
\def \dsR {\text{$\mathds{R}$}}

\DeclareMathOperator{\Var}{Var}

\DeclareMathOperator{\diag}{diag}

\DeclareMathOperator{\ND}{N}
\DeclareMathOperator{\LND}{LN}

    \def \mI {\text{\boldmath$I$}}

\def \uvec {\text{\boldmath$u$}}    
    
\def \wvec {\text{\boldmath$w$}}    
\def \xvec {\text{\boldmath$x$}}    
\def \yvec {\text{\boldmath$y$}}    \def \mY {\text{\boldmath$Y$}}
\def \zvec {\text{\boldmath$z$}}    \def \mZ {\text{\boldmath$Z$}}

\def \Xg         {\text{$\bm{X}_{\gamma}$}}
\def \Xgi         {\text{$\bm{x}_{\gamma,i}$}}

\def \qg         {\text{$q_{\gamma}$}}

\def \bg         {\text{\boldmath$\beta$\unboldmath$_{\gamma}$}}

\def \alphavec        {\text{\boldmath$\alpha$}}
\def \betavec         {\text{\boldmath$\beta$}}
\def \gammavec        {\text{\boldmath$\gamma$}}

\def \varepsilonvec   {\text{\boldmath$\varepsilon$}}

\def \mSigma   {\mathbf{\Sigma}}

\def \mOmega   {\mathbf{\Omega}}

\def \nullvec {\mathbf{0}}
\def \onevec {\mathbf{1}}

\newtheorem{defin}{Definition} 
\newtheorem{prop}{Proposition}

\newcounter{mynotation}

\usepackage{paralist}

\renewenvironment{itemize}[1]{\begin{compactitem}#1}{\end{compactitem}}

\makeatletter
\def\@seccntformat#1{\@ifundefined{#1@cntformat}%
	{\csname the#1\endcsname\quad}  
	{\csname #1@cntformat\endcsname}
}
\let\oldappendix\appendix 
\renewcommand\appendix{%
	\oldappendix
	\newcommand{\section@cntformat}{\appendixname~\thesection\quad}
}
\makeatother

\usepackage{scalerel,stackengine}
\stackMath
\newcommand\reallywidehat[1]{%
\savestack{\tmpbox}{\stretchto{%
  \scaleto{%
    \scalerel*[\widthof{\ensuremath{#1}}]{\kern-.6pt\bigwedge\kern-.6pt}%
    {\rule[-\textheight/2]{1ex}{\textheight}}
  }{\textheight}%
}{0.5ex}}%
\stackon[1pt]{#1}{\tmpbox}%
}

\begin{document}
\begin{titlepage}
\title[Bayesian Variable Selection for Non-Gaussian Responses]{Bayesian Variable Selection for Non-Gaussian Responses: A Marginally-Calibrated Copula Approach}
\author{Nadja Klein\emailx{nadja.klein@hu-berlin.de}\\
	School of Business and Economics, Humboldt-Universit\"at zu Berlin, Berlin, Germany,
	\and
	Michael Stanley Smith\emailx{mike.smith@mbs.edu} \\
	Melbourne Business School, University of Melbourne, 200 Leicester Street, Carlton, VIC 3053, Australia.}
\begin{abstract}
We propose a new highly flexible and tractable Bayesian approach to undertake variable selection in non-Gaussian regression models.
It uses a copula decomposition for the joint distribution of observations on the dependent variable. This allows 
the marginal distribution of the dependent variable to be calibrated accurately using a nonparametric or other
estimator.
The family of copulas employed are `implicit
copulas' that are constructed from existing hierarchical Bayesian models widely used for variable selection,
and we establish some of their properties.
Even though the copulas are high-dimensional, they can be estimated efficiently and quickly using 
Markov chain Monte Carlo (MCMC). A simulation study shows that when the responses are non-Gaussian
the approach selects variables more accurately than contemporary benchmarks.
A real data example in the Web Appendix illustrates that accounting for even mild deviations from normality
can lead to a substantial increase in accuracy.
To illustrate the full potential of our approach we extend it to spatial variable selection
for fMRI. Using real data, we show our method allows for voxel-specific marginal calibration of the magnetic resonance signal
at over 6,000 voxels, leading to an increase in the quality of the activation maps.
\end{abstract} 
\end{titlepage}
\maketitle
\section{Introduction}
Bayesian approaches to selecting covariates in regression models are well established; 
see~\cite{HarSil2009} and~\cite{bottolo2010} for overviews. However, most work
remains focused on Gaussian regression models, and extensions to the non-Gaussian case are
 limited. In particular, the importance of `marginal calibration' in
Bayesian variable (i.e. covariate) selection (BVS) is unexplored.
Following~\cite{Gneetal2007}, by marginal calibration we mean 
the statistical consistency between the unconditional
probability distribution of the dependent
variable and its observations, with a more
formal definition given by~\citet[Defn. 2.6a]{gneiting2013}.
We address this  
here by proposing a new  approach to BVS that is based on a
copula decomposition for a vector of observations of length $n$ on the dependent variable.
The impact of the covariates on the dependent variable is captured by the copula function only.
This separates the
task of selecting covariates from that of
modeling the marginal distribution of the dependent variable; the latter of which can then be
calibrated accurately. 
For the copula function we propose a new family of `implicit copulas', 
which are constructed from existing popular Bayesian hierarchical regression
models used for selecting covariates. 
By an implicit copula we mean the copula that is implicit
in a multivariate distribution and obtained by inverting the usual expression of Sklar's theorem as in~Sec.~3.1 of~\citet{Nel2006}.
The result is a general and tractable approach that extends BVS
to have an accurately calibrated margin for the dependent variable.

Low dimensional copulas are often used to capture dependence between multiple variables. 
Here the copula is used in a
different way to capture the dependence between multiple observations on one dependent
variable. The specification of this $n$-dimensional copula is the key ingredient
of our method. To do so we consider a Gaussian linear model for $n$ observations on
another dependent variable, which we call a `pseudo-response' because it is not observed directly. Gaussian
spike-and-slab priors with selection indicator variables $\gammavec$
are employed for the coefficients. Integrating out these coefficients gives a Gaussian distribution
for the pseudo-response vector conditional on the covariates and $\gammavec$, and its implicit copula is a Gaussian copula~\citep{Song2000}
with parameter matrix that is a function of the covariate values and $\gammavec$.
Finally, to obtain our copula family we mix this Gaussian copula over the scaling factor $g$ of the
non-zero coefficients with respect to 
the different hyper-priors suggested by~\cite{LiaPauMolClyBer2008}. The resulting implicit copulas are 
mixtures of Gaussian copulas.

Because of its high dimension, it is difficult to evaluate our copula family directly. However, we show how to
construct an MCMC sampler to undertake stochastic search variable selection~\citep{GeoMcC1993}, where the scaling factor $g$ is sampled using the Hamiltonian Monte Carlo (HMC)
method
of~\cite{HofGel2014}. Careful use of matrix identities for the computations makes application 
of the method to high dimensions practical. A simulation study compares
our approach to Gaussian BVS and the method of~\cite{RosRub2017}. It shows that accurate
marginal
calibration of the distribution the dependent variable---an intrinsic feature of our copula model---can 
result in more accurate covariate selection and predictive densities.

However, the main application of our approach is to spatial variable selection for functional magnetic resonance imaging (fMRI). 
In these studies a large vector of  binary indicators signifies  
which voxels in a partition of the brain are active, so that it is 
large variable selection problem. We follow~\cite{SmiPueAueFah2003}, \cite{SmiFah2007},
\cite{LiZhang2010} and~\cite{goldsmith2014}
and use an Ising model as a prior to smooth the binary indicators spatially, but employ our copula model to allow 
for voxel-wise marginal calibration 
of the magnetic resonance (MR) signal. The neuroimaging literature suggests that
accounting for such deviations from normality in the MR signal is important to obtain accurate activation maps~\citep{eklund2017}. Application of our approach to data from a visual experiment with $6192$ voxels
shows this to be true here, and also produces much more accurate voxel-wise
predictive distributions for the MR signal, as measured by the logarithmic scores. Importantly,
the approach is both fast to implement and can be readily generalized, including to 
other binary random field priors for spatial smoothing of the binary indicators.
 
A number of other Bayesian approaches consider
variable selection for non-Gaussian continuous-valued data.
These include
conditionally Gaussian models, where the disturbances follow a mixture of normals and/or
data transformations of the dependent variable are
considered, as in~\cite{SmiKoh1996} and~\cite{GotRaf2007}.
 \citet{RosRub2017} propose a BVS approach that allows for skewness and heavy tails by employing two-piece
 Gaussian and Laplace distributions for the errors.
 \cite{ChuDun2009} consider variable selection for a distributional regression model constructed through
 a probit stick-breaking process,
 \cite{Kundu2014} consider selection when the errors are modelled non-parametrically.  
\citet{YuCheReeDun2013} propose variable selection in Bayesian quantile regression.
However, none of these approaches fits our copula framework, nor
ensure accurate calibration of the marginal
distribution of the response. 
\cite{ShaSou2018} proposed an alternative class of priors for 
regression coefficients based on copulas, and \cite{kraus2017} used a D-vine copula to capture flexibly
the dependence between covariates and response in a regression model.
However both use copulas in a very different way than suggested here
and do not generalize existing BVS schemes as our approach does. Last,~\cite{KleSmi2019}
construct implicit copulas from regularized smoothers, and our copula family
extends these to variable selection.

The rest of this paper is structured as follows. Section~\ref{sec:implicit:copula} outlines our approach,
including our proposed copula family. Section~\ref{sec:estimation}
details how to compute Bayesian inference, including the predictive densities
and Bayes factors.
Section~\ref{sec:sim} contains the simulation study,
Section~\ref{sec:fmri} extends the approach to spatial variable selection for fMRI data, and Section~\ref{sec:discussion} concludes. A Web Appendix contains extensive additional material,
including proofs, copula properties, details of the estimation algorithms, and the
in-depth analysis of an 
additional regression dataset with $p=252$ correlated covariates.

\setlength{\abovedisplayskip}{1pt}
\setlength{\belowdisplayskip}{1pt}
\section{Variable Selection in Regression using Copulas}\label{sec:implicit:copula}

\subsection{Marginally calibrated variable selection} \label{subsec:margcal}
Consider a 
vector $\bm{Y}=(Y_1,\ldots,Y_n)'$ of $n$ realisations on a continuous
dependent variable, along with an $n\times p$ design matrix $\bm{X}$
for $p$ regression covariates.
Bayesian approaches to variable selection 
usually proceed by introducing a vector of binary
indicator variables $\gammavec=(\gamma_1,\ldots,\gamma_p)'$, such that the $j$th covariate
is included in the regression if $\gamma_j=1$, and excluded if $\gamma_j=0$. 
When the dependent variable is non-Gaussian, the most
common approach is to consider non-Gaussian distributions for the disturbance to 
a linear model;
see~\cite{Kundu2014}, \cite{RosRub2017} and references therein. 
Thus, 
a non-Gaussian distribution is selected for $Y_i$, conditional on $\bm{X}$ and $\gammavec$.
In this paper we suggest an alternative approach based on copulas
that allows
the marginal distribution of $Y_i$, unconditional on $\bm{X}$ and $\gammavec$, to be chosen.
We model the joint density of $\bm{Y}|\bm{X},\gammavec$  as
\begin{equation}\label{eq:copmod}
p(\yvec|\bm{X},\gammavec)=c_{\mbox{\tiny BVS}}(\uvec|\bm{X},\gammavec)\prod_{i=1}^n p_{Y}(y_i)\,,
\end{equation}
where $\yvec=(y_1,\ldots,y_n)'$, $\uvec=(u_1,\ldots,u_n)'$, $u_i=F_{Y}(y_i)$ and 
 the distribution of $Y_i$ is assumed to be {\em marginally} invariant with respect to $\bm{X}$ and $\gammavec$, with density $p_Y$ and
 distribution function $F_Y$. The impact of 
the covariate values $\bm{X}$ and model indicators $\gammavec$ on $\bm{Y}$ {\em jointly} is captured through
the copula with density $c_{\mbox{\tiny BVS}}$, which is a function of $\bm{X}$
and $\gammavec$. For this
we use the copula proposed in Section~\ref{subsec:gen:idea} below.

A major advantage of employing~(\ref{eq:copmod}) is that it separates the modeling of the marginal
distribution $F_Y$ of the data, from the task of selecting the covariates. Therefore,
$F_Y$ can be calibrated accurately, 
and we model it non-parametrically in our work. Variable selection
is based on the posterior distribution
of $\gammavec$, which is given by 
\begin{equation}
p(\gammavec|\bm{X},\yvec)\propto p(\yvec|\bm{X},\gammavec)p(\gammavec) \propto 
c_{\mbox{\tiny BVS}}(\uvec|\bm{X},\gammavec)p(\gammavec)\,,
\label{eq:gammapost}
\end{equation}
with model prior $p(\gammavec)$. 
A major aim of this paper is to show that adopting~\eqref{eq:copmod} with our proposed copula 
provides a very general, but tractable, approach to undertaking variable selection and model
averaging for non-Gaussian regression data. 
  
We make three
remarks concerning the appropriateness of the
decomposition at~\eqref{eq:copmod}. 
First, regression models are usually specified conditional on parameters
for the mean, variance
and possibly other moments. 
In contrast, the expressions 
above are unconditional on such parameters. We show in Part~B of the Web Appendix
that when also
conditioning on
additional model parameters introduced below in Section~\ref{subsec:gen:idea}, 
the distribution of $Y_i$ is a function of the 
covariates, as is expected in a regression model. 
Second, we show in Part~A of the Web Appendix
that in a Gaussian linear regression model with a
zero mean g-prior for the regression coefficients, the margin of $Y_i$
with the coefficients and error variance 
integrated out,
is asymptotically independent of $\bm{X}$.
Third, the predictive density 
arising from the copula model at~\eqref{eq:copmod} is a function of the covariate values and indicator variables $\gammavec$. 
To see this, consider
a new realization $Y_{n+1}$, with a $p\times 1$  vector of covariate values $\xvec_{n+1}$. 
Let $\bm{X}^+=[\bm{X}'|\xvec_{n+1}]'$ and $\uvec^+=(\uvec',F_Y(y_{n+1}))'$, then 
from~\eqref{eq:copmod}, $Y_{n+1}$ has predictive
density 
\begin{equation}
p(y_{n+1}|\bm{X}^+,\gammavec,\yvec)=\frac{p(y_{n+1},\yvec|\bm{X}^+,\gammavec)}{p(\yvec|\bm{X},\gammavec)}
=\frac{c_{\mbox{\tiny BVS}}(\uvec^+|\bm{X}^+,\gammavec)}{c_{\mbox{\tiny BVS}}(\uvec|\bm{X},\gammavec)} p_Y(y_{n+1})\,.
\label{eq:pred1}
\end{equation}
This is a function of the observed values of all the covariates and $\gammavec$. 
Moreover, marginalizing over the posterior of $\gammavec$ gives
the posterior predictive density for $Y_{n+1}|\bm{X}^+,\yvec$ as 
\begin{equation}
p(y_{n+1}|\bm{X}^+,\yvec)=\sum_\gammavec p(y_{n+1}|\bm{X}^+,\gammavec,\yvec) p(\gammavec|\bm{X},\yvec)\,.
\label{eq:pred2}
\end{equation}
This forms the basis of the predictive
distribution of $Y_{n+1}$ from the copula model, and we give a 
computationally tractable expression for~\eqref{eq:pred2}
in Section~\ref{subsec:pred}. 

\subsection{Variable selection copula}\label{subsec:gen:idea}
Key to our approach is the specification
of our proposed copula  with density 
$c_{\mbox{\tiny BVS}}$. 
To derive this, consider the linear model
\begin{equation}
{\widetilde\mZ}= \bm{X}_{\gamma}\betavec_{\gamma}+\varepsilonvec\,,
\label{eq:regression}
\end{equation}
where $\widetilde{\mZ}=(\tilde Z_1,\ldots,\tilde Z_n)'$, 
$\bm{X}_\gamma$ is an $n\times q_\gamma$ sub-matrix of $\bm{X}$ that comprises the columns of
$\bm{X}$ where $\gamma_j=1$ (so $q_\gamma=\sum_{j=1}^p \gamma_j$), $\betavec_\gamma$ are the corresponding regression coefficients, 
and   
$\varepsilonvec\sim\ND(\nullvec,\sigma^2\mI)$. We
refer to $\widetilde{\bm{Z}}$ as a vector of observations
on a `pseudo-response', 
because it is not observed
directly in our model. 
Following~\cite{SmiKoh1996,GeoMcC1997,LiaPauMolClyBer2008}
and many others, the conjugate g-prior 
$\bg|\bm{X},\sigma^2,\gammavec,g\sim\ND(\bm{0},g\sigma^2(\Xg'\Xg)^{-1})$, $g>0$,
is used for the non-zero coefficients.
Its conjugacy and scaling prove attractive features for constructing 
the variable selection copula.

To construct the variable selection copula, 
we first extract the implicit copula of the distribution of $\widetilde{\bm{Z}}$ conditional on $\bm{X},\gammavec,g,\sigma^2$, but
with $\betavec_\gamma$ integrated out.
This is
$\widetilde\mZ| \bm{X},\gammavec,g,\sigma^2\sim N(\bm{0};\bf{\Omega})$,
where
\[
\bm{\Omega}=\sigma^2\left(\mI-\tfrac{g}{1+g}\Xg(\Xg'\Xg)^{-1}\Xg'\right)^{-1}=\sigma^2(\mI+g\Xg(\Xg'\Xg)^{-1}\Xg')\,,
\]
which follows from integrating out $\betavec_\gamma$ as a normal,
and applying the Woodbury formula. The implicit copula is the Gaussian copula~\citep{Song2000}, with parameter
matrix equal to the correlation $\bm{R}$ of the distribution, and density
$
c_{\mbox{\tiny Ga}}(\uvec;\bm{R})=|\bm{R}|^{-1/2}\exp\left(-\frac{1}{2}\zvec'(\bm{R}^{-1}-\mI)\zvec  \right)\,,
$
where $\zvec=(\Phi^{-1}_1(u_1),\ldots,\Phi^{-1}_1(u_n))'$ and $\Phi_1$ is the standard normal distribution function. 
To derive $\bm{R}$ we standardize $\mOmega$ by the variances
\[
\Var(\tilde Z_i| \bm{X},\gammavec,g,\sigma^2)=\sigma^2(1+g\Xgi'(\Xg'\Xg)^{-1}\Xgi)\equiv\sigma^2 s_i^{-2}\,,
\mbox{ for}\,\, i=1,\ldots,n\,,
\]
where $\Xgi'$ is the $i$th row of $\bm{X}_\gamma$. Let $\bm{S}_\gamma\equiv \bm{S}(\bm{X},\gammavec,g)=\diag(s_1,\ldots,s_n)$, then
\begin{equation}\label{eq:correlationmatrix}
\bm{R}\equiv \bm{R}(\bm{X},\gammavec,g)= \frac{1}{\sigma^2}\bm{S}_\gamma \mOmega \bm{S}_\gamma = \bm{S}_\gamma\left(\mI+g\Xg(\Xg'\Xg)^{-1}\Xg'\right)\bm{S}_\gamma\,.
\end{equation}
Note that the location and scale of $\widetilde{\bm{Z}}$ are unidentified in its copula, 
and $\bm{R}$ is not a function of  $\sigma^2$. Therefore,
without loss of generality,
we set $\sigma^2=1$ and do not include an intercept in $\bm{X}$. We stress here that this does not mean the observational data $Y_i$ has zero mean or fixed scale, which is instead captured through $F_Y$ in~\eqref{eq:copmod}.
\begin{sidewaystable}
	\caption{Summary of the one-to-one transformations between the dependent variable $Y_i$, 
		copula variable $U_i$ specified in Section~\ref{sec:implicit:copula}, and the standardized pseudo-response
		$Z_i=\frac{s_i}{\sigma}\tilde{Z}_i$
		specified in Section~\ref{subsec:posteval}. Also given are 
		the joint densities of $\bm{Y}=(Y_1,\ldots,Y_n)'$, $\bm{U}=(U_1,\ldots,U_n)'$
		and $\bm{Z}=(Z_1,\ldots,Z_n)'$ conditioning on $\bm{X},\gammavec$ and with/without
		$g$. 
		Above, $\yvec=(y_1,\ldots,y_n)'$, $\uvec=(u_1,\ldots,u_n)'$, $\zvec=(z_1,\ldots,z_n)'$ and the $n\times n$ correlation matrix $\bm{R}\equiv \bm{R}(\bm{X},\gammavec,g)$ is specified at~\eqref{eq:correlationmatrix}.}\label{tab:model}
	\begin{tabular}{lccc}
		\hline
		\hline
		& Observed Data & Copula Data & Standardized Pseudo-data\\
		\hline
		Variable & $Y_i$ & $U_i=F_Y(Y_i)$ & $Z_i=\Phi_1^{-1}(U_i)$\\
		Domain & $\dsR$ & $[0,1]$ & $\dsR$\\
		Marginal distribution &$F_Y$ &Uniform &Standard Normal\\
		Joint density  &
		$p(\yvec|\bm{X},\gammavec,g)=\phi(\zvec;\nullvec,\bm{R})$ & $p(\uvec|\bm{X},\gammavec,g)=c_{\mbox{\tiny{Ga}}}(\uvec|\bm{X},\gammavec,g)$ & $p(\zvec|\bm{X},\gammavec,g)=\phi(\zvec;\nullvec,\bm{R})$\\
		conditional on $\bm{X},\gammavec,g$ &$\times \prod_{i=1}^n \frac{p_Y(y_i)}{\phi_1(z_i)}$ & & \\
		Joint density  & $p(\yvec|\bm{X},\gammavec)=c_{\mbox{\tiny{BVS}}}(\uvec|\bm{X},\gammavec)$ & $p(\uvec|\bm{X},\gammavec)=c_{\mbox{\tiny{BVS}}}(\uvec|\bm{X},\gammavec)$ & $p(\zvec|\bm{X},\gammavec)=\int\phi(\zvec;\nullvec,\bm{R})p(g)\mathrm{d}g$\\
		conditional on $\bm{X},\gammavec$ &$\times \prod_{i=1}^n p_Y(y_i)$&&\\
		\hline\hline
	\end{tabular}
\end{sidewaystable}

Finally, 
we mix over $g$ with respect to its prior $p(g)$ 
to obtain the variable selection copula as a continuous
mixture of Gaussian copulas, as defined below.
\begin{defin}
\label{defin:cbvs}
Let $C_{\mbox{\tiny Ga}}$ and $c_{\mbox{\tiny Ga}}$ be 
the Gaussian copula function and density, respectively. Then if
$\widetilde{\bm{Z}}$ follows the linear model at~(\ref{eq:regression}), with the g-prior for $\bm{\beta}_\gamma$, and $p(g)$ is a proper density for $g>0$,
then we call
$C_{\mbox{\tiny BVS}}(\bm{u}|\bm{X},\gammavec)=\int C_{\mbox{\tiny Ga}}(\bm{u};\bm{R}(\bm{X},\gammavec,g))p(g)
\mathrm{d}g$
a variable selection copula, with density function
\[
c_{\mbox{\tiny BVS}}(\bm{u}|\bm{X},\gammavec)=\int c_{\mbox{\tiny Ga}}(\bm{u};\bm{R}(\bm{X},\gammavec,g))p(g)
\mathrm{d}g\,.
\] 
It is straightforward to show 
the function 
$C_{\mbox{\tiny BVS}}(\bm{u}|\bm{X},\gammavec)$
is a well-defined copula function.
\end{defin}
\noindent 
Table~\ref{tab:model} depicts the transformations underlying the construction of $C_{\mbox{\tiny BVS}}$. 
Part~C of the Web Appendix gives some properties of $C_{\mbox{\tiny BVS}}$.
We consider the three priors discussed by~\citet{LiaPauMolClyBer2008} for $g$, and a 
point mass, as listed below:
\begin{itemize}
\item[(a)] \emph{Hyper-$g$ prior:} with density $p(g)=\tfrac{a-2}{2}(1+g)^{-a/2}$, which is proper for $a>2$. 
This implies a beta prior on the shrinkage factor $g/(1+g)\sim\mbox{Beta}(1,0.5a-1)$, and we set $a=4$ leading to a uniform prior on this factor. 
\item[(b)] \emph{Hyper-$g/n$ prior:} with density $p(g)=\tfrac{a-2}{2n}(1+g/n)^{-a/2}$ and $a=4$. 
\item[(c)] \emph{Zellner-Siow prior:} with density $p(g)=\tfrac{\sqrt{n/2}}{\Gamma(1/2)}c^{-3/2}\exp(-n/(2g))$. 
\item[(d)] \emph{Point mass prior:} We also consider fixing $g=100$ and $g=n$
for comparison.
\end{itemize}

While computing the integral over $g$ in Defn.~\ref{defin:cbvs} is possible using numerical methods,
in general it 
is difficult to evaluate  $C_{\mbox{\tiny BVS}}$ or  $c_{\mbox{\tiny BVS}}$ directly
because $\bm{R}(\bm{X},\gammavec,g)$ is an $n$-dimensional matrix.
Instead, 
we generate $g$ as part of an MCMC scheme, as discussed in 
Section~\ref{sec:estimation}. 
We use the popular prior
$p(\gammavec)=B(p-q_{\gamma}+1,q_{\gamma}+1)$, where $B$ is the 
beta function, which
implies $p(q_{\gamma})=1/(p+1)$. 
\section{Estimation and Inference}\label{sec:estimation}
Estimation of~(\ref{eq:copmod}) requires estimation
of both the marginal $F_Y$ and copula parameters $\gammavec$. 
It is common to use two-stage estimators, where $F_Y$ is estimated first, followed by 
$\gammavec$, because they are much faster and often involve only a 
minor loss of efficiency~\citep{joe2005}.
\cite{GraLis2017} and~\cite{KleSmi2019} integrate out uncertainty for $F_Y$ using a Bayesian non-parametric estimator, but find that this
does not improve the accuracy of inference
meaningfully, as we also demonstrate in an empirical example in Part~F of the Web Appendix.
Therefore, we adopt a two-stage estimator, and use the
adaptive kernel density estimator (KDE) of~\cite{shimazaki2010} to estimate $F_Y$.

\subsection{Posterior evaluation}\label{subsec:posteval}
We follow~\cite{GeoMcC1993} and evaluate the posterior
of $\gammavec$ using MCMC. However,  
direct computation of the posterior mass at~(\ref{eq:gammapost}) 
is slow because computing 
$c_{\mbox{\tiny BVS}}$ requires integration over $g$, so we generate $g$ as part of
the MCMC scheme. To implement the sampler we follow \cite{KleSmi2019} and express the likelihood conditional on $g$ in closed form by transforming to the (normalized) pseudo-response as follows. 
Let $\bm{Z}=(Z_1,\ldots,Z_n)'=\frac{1}{\sigma}\bm{S}_\gamma\widetilde{\bm{Z}}$, where $\widetilde{\bm{Z}}$ is the pseudo-response at~\eqref{eq:regression}, then it follows from Section~\ref{subsec:gen:idea} that
$\bm{Z}|\bm{X},\gammavec,g\sim N(\bm{0},\bm{R}(\bm{X},\gammavec,g))$. 
Moreover, $Y_i$ can be 
expressed in terms of $Z_i$ as 
$Y_i=F_Y^{-1}(\Phi_1(Z_i))$, so that by a change of variables from $\bm{Y}$ to $\bm{Z}$,   
\begin{equation}
p(\yvec|\bm{X},\gammavec,g)
= p(\zvec|\bm{X},\gammavec,g)\prod_{i=1}^n\frac{p_{Y}(y_i)}{\phi_1(z_i)}  
 = \phi(\zvec;\nullvec,\bm{R})\prod_{i=1}^n\frac{p_{Y}(y_i)}{\phi_1(z_i)}
 \,,\label{eq:likelihood:y:cond:g}
\end{equation}
where the Jacobian of the transformation is $|\frac{d\bm{z}}{d\bm{y}}|=\prod_{i=1}^n \frac{p_Y(y_i)}{\phi_1(z_i)}$, $\phi_1$ is the standard normal density, and
$\phi(\zvec;\bm{0},\bm{R})$ is the density of a $N(\bm{0},\bm{R})$ distribution.

While all the terms in the right-hand side of~\eqref{eq:likelihood:y:cond:g}
are known, the $n \times n$ matrix $\bm{R}$ cannot be computed directly for large $n$. 
To evaluate the posterior, we employ the following sampler.

\noindent {\bf MCMC Sampler}\\
\noindent At each sweep:\\
\noindent \underline{Step~1.} Randomly partition $\gammavec$ into pairs of elements.\\
\noindent \underline{Step~2.} For each pair $(\gamma_i,\gamma_j)$, generate from $p(\gamma_i,\gamma_j|\{\gammavec\backslash \gamma_i,\gamma_j\},\bm{X},g,\yvec)$.\\
\noindent \underline{Step~3.} Generate from $p(g|\bm{X},\gammavec,\yvec)$ using Hamiltonian Monte Carlo.

In forming the partition in Step~1, if $p$ is odd-valued one element is simply selected twice, so that pairs of elements $(\gamma_i,\gamma_j)$ are always generated in Step~2. Sampling pairs of elements of $\gammavec$ in random order
helps to improve mixing of the chain. To implement Step~2,
from~(\ref{eq:likelihood:y:cond:g}) the joint posterior of the indicators is
\begin{eqnarray*}
p(\bm{\gamma}|\bm{X},g,\bm{y}) &\propto &p(\yvec|\bm{X},\gammavec,g)p(\bm{\gamma})
\propto \phi(\zvec;\bm{0},\bm{R}(\bm{X},\gammavec,g)) p(\bm{\gamma}) \nonumber\\
&\propto &|\bm{R}(\bm{X},\gammavec,g)|^{-1/2} \exp\left\{ -\frac{1}{2} \left( \zvec'\bm{R}(\bm{X},\gammavec,g)^{-1}\zvec
	\right) \right\}p(\gammavec)\equiv A(\gamma_i,\gamma_j)\,. \label{eq:cposbvs}
\end{eqnarray*}
Simulating $(\gamma_i,\gamma_j)$ involves
computing $A$ for the four possible configurations \\${\cal S}\equiv \{(0,0),
(0,1), (1,0), (1,1) \}$, and then setting
\begin{equation}
p((\gamma_i,\gamma_j)|\{\gammavec\backslash (\gamma_i,\gamma_j)\},\bm{X},g,\yvec)=\frac{1}{1+h}\,,
\;\;
h=\sum_{( \tilde \gamma_i, \tilde \gamma_j)\in \{{\cal S}\backslash (\gamma_i,\gamma_j)\}} \frac{A(\tilde\gamma_i, \tilde \gamma_j)}{A(\gamma_i,\gamma_j)}\,,
\label{eq:cposbvs2}
\end{equation}
where all ratios are computed on the logarithmic scale.
Fast computation of $A$ for the four configurations in ${\cal S}$ can be done
using a number of matrix identities; see Part~D of the Web Appendix,
where at no stage is the full $n\times n$ matrix $\bm{R}$ 
computed directly. 

Generating $g$ at Step~3 uses
an HMC step for $\tilde g=\log(g)$. We use a variant of the leapfrog integrator of~\cite{Nea2011} with the dual averaging approach of~\citet{HofGel2014}. This requires computation
of $\log(p(\tilde g|\bm{X},\gammavec,\yvec))$ up to an additive constant, and its derivative. 
Part~D of the Web Appendix derives analytical expressions for these, and outlines the HMC 
step in greater detail. We found that the sampler above works well in our applications, although a referee highlighted that such samplers may mix poorly for large $p$,
in which case adaptive non-Markov samplers as in~\cite{griffin2017search} may be preferable.

\subsection{Inference}\label{subsec:inference}
The sampler above produces Monte Carlo draws $\{\gammavec^{[k]},g^{[k]}; k=1,\ldots,K\}$ from the posterior
 $p(\gammavec,g|\bm{X},\yvec)$ and is used to compute posterior estimates as detailed below.
\subsubsection{Variable selection}
Variables can be selected using the marginal posteriors, which are estimated
as
\[
\mbox{Pr}(\gamma_i=1|\bm{X},\yvec)\approx 
\frac{1}{K}\sum_{k=1}^K
 \mbox{Pr}(\gamma_i=1|\{\gammavec^{[k]}\backslash(\gamma_i,\gamma_j)\},\bm{X},g^{[k]},\yvec)\,.
 \]
 To evaluate
 the term in this summation, at Step~2 of the sampler for the single pair $(\gamma_i,\gamma_j)$
 that contains $\gamma_i$, the following is computed
\[
\mbox{Pr}(\gamma_i=1|\{\gammavec\backslash(\gamma_i,\gamma_j)\},\bm{X},g,\yvec)=
\frac{A(1,0)+A(1,1)}{A(0,0)+A(1,0)+A(0,1)+A(1,1)} \,,
\]
where the four values of the bivariate function $A(\gamma_i,\gamma_j)$ are already computed at~\eqref{eq:cposbvs2}.
\subsubsection{Predictive density}\label{subsec:pred}
In general,
direct evaluation of the
predictive density of a new observation $Y_{n+1}$ with covariates $\xvec_{n+1}$
at~(\ref{eq:pred1}) is infeasible because evaluating $c_{\mbox{\tiny BVS}}$
is also. However, 
the posterior predictive density at~(\ref{eq:pred2}) can still be evaluated as:
\begin{equation}\label{eq:approx}
p(y_{n+1}|\bm{X}^+,\yvec) = \sum_{\gammavec} \int \int
p(y_{n+1}|\bm{X}^+,\betavec_{\gamma},\gammavec,g,\yvec)
p(\betavec_\gamma,\gammavec,g|\bm{X},\bm{y})\mathrm{d}(\betavec_\gamma,g)\,.
\end{equation}
The predictive density inside the integrals above can obtained by considering a change of variables
from $Y_{n+1}$ to $Z_{n+1}=\Phi_1^{-1}(F_Y(Y_{n+1}))$,
with Jacobian $\frac{p_Y(y_{n+1})}{\phi_1(z_{n+1})}$, as 
\[
p(y_{n+1}|\bm{X}^+,\betavec_\gamma,\gammavec,g,\yvec) = p(z_{n+1}|\bm{X}^+,\betavec_\gamma,g,\gammavec,\zvec)
\frac{p_Y(y_{n+1})}{\phi_1(z_{n+1})}\,.
\]
From~\eqref{eq:regression},
$\tilde Z_{n+1}|\xvec_{\gamma,i},\betavec_\gamma,\gammavec,g\sim N(\xvec_{\gamma,{n+1}}'\betavec_\gamma,\sigma^2)$
independently when conditioning on $\betavec_\gamma$ (whereas 
the elements of $\widetilde{\bm Z}$ are dependent unconditional on $\betavec_\gamma$).
Then, because $Z_{n+1}=\frac{s_{n+1}}{\sigma} \tilde Z_{n+1}$, 
\begin{equation}\label{eq:cbetapred}
p(y_{n+1}|\bm{X}^+,\betavec_\gamma,\gammavec,g,\yvec) = \frac{1}{s_{n+1}} \phi_1\left(\frac{z_{n+1}-s_{n+1}\xvec_{\gamma,n+1}' \betavec_{\gamma}}{s_{n+1}}\right)\frac{p_{Y}(y_{n+1})}{\phi_1(z_{n+1})}\,,
\end{equation}
where $s_{n+1}=(1+g \xvec_{\gamma,n+1}'(\Xg'\Xg)^{-1}\xvec_{\gamma,n+1})^{-1}$, and $\xvec_{\gamma,n+1}$ are
the elements of $\xvec_{n+1}$ that correspond to $\gammavec$. Notice that 
$\sigma^2$ cancels out in the above because it is 
unidentified in the copula, and plays no role in the predictions.

An expression for the 
posterior predictive density is obtained by plugging~\eqref{eq:cbetapred} into~\eqref{eq:approx}.
The integrals and summation can be evaluated in the usual Bayesian fashion
by averaging over Monte Carlo iterates from the posterior $p(\betavec_\gamma,\gammavec,g|\bm{X},\bm{y})$.
However, this
requires the additional generation of $\betavec_\gamma$ at each sweep of the sampler.
A faster approximation that avoids this---and which we have found to 
be almost as accurate empirically---is to plug in the posterior
expectation of $\betavec_{\gamma}$ conditional on $\gammavec,g$, given by 
$\widehat{\betavec}_\gamma
=\frac{g}{1+g}(\Xg'\Xg)^{-1}\Xg' \bm{S}_{\gamma}^{-1}\zvec$.
The main components required for the evaluation of $\widehat{\betavec}_\gamma$ are
computed previously at each sweep of the sampler. 
Thus, a fast and accurate predictive density estimator
can be constructed using the $K$ Monte Carlo iterates from the 
sampler as
\begin{equation}
\hat{p}(y_{n+1}|\bm{X}^+,\yvec)=
\frac{\hat{p}_Y(y_{n+1})}{\phi_1(\Phi_1^{-1}(\hat{F}_Y(y_{n+1})))}
\left\{
\frac{1}{K}\sum_{k=1}^K
\frac{1}{s_{n+1}^{[k]}}\phi_1\left( 
	\frac{\Phi_1^{-1}(\hat{F}_Y(y_{n+1}))-s_{n+1}^{[k]}\xvec_{\gamma^{[k]},n+1}'\widehat{\betavec}_{\gamma^{[k]}}}{s_{n+1}^{[k]}}
\right)
\right\}
\,,\label{eq:phatyx}
\end{equation}
with 
$s_{n+1}^{[k]}\equiv (1+g^{[k]}\xvec_{\gamma^{[k]},n+1}'(\bm{X}_{\gamma^{[k]}}'\bm{X}_{\gamma^{[k]}})^{-1}\xvec_{\gamma^{[k]},n+1})^{-1}$.  Last, the mean of~\eqref{eq:approx} is the regression function estimator, as outlined 
in Part~D of the Web Appendix.

\subsubsection{Bayes factor}
We derive the  Bayes factor for a pair of covariate subsets $\gammavec$ and $\widetilde\gammavec$.
For $\gammavec$,  let  $\bm{U}_{\gamma}$ be an upper triangular Cholesky factor, such that $\bm{U}_{\gamma}'\bm{U}_{\gamma}=\bm{X}_\gamma'\bm{X}_\gamma$ and $\bm{M}_{\gamma}=\bm{X}_\gamma \bm{U}_{\gamma}^{-1}$. 
Then if $\bm{M}_{\gamma}=\{m_{\gamma,ij}\}$, we can express $s_{i}=(1+g\sum_{j=1}^{q_\gamma} m_{\gamma,ij}^2)^{-1/2}$ for $i=1,\ldots,n$,  and 
$\bm{R}(\bm{X},\gammavec,g)^{-1}=\bm{S}_{\gamma}^{-1}(\mI-\frac{g}{1+g}\bm{M}_{\gamma}\bm{M}_{\gamma}')\bm{S}_{\gamma}^{-1}$; with similar expressions for $\widetilde\gammavec$

\begin{prop}\label{prop:BF} 
The  Bayes factor for model $\gammavec$ over model $\widetilde\gammavec$ is given by
\begin{equation*}
\begin{aligned}
\mbox{BF}(\gammavec | \widetilde\gammavec)&=
\int_0^\infty\prod_{i=1}^n\left\lbrack(1+g\sum_{j=1}^{\qg}m_{\gamma,ij}^2)^{\frac{1}{2}}(1+g\sum_{j=1}^{q_{\tilde\gamma}}m_{\tilde\gamma,ij}^2)^{\frac{1}{2}}\right\rbrack(1+g)^{-\frac{\qg-q_{\tilde\gamma}}{2}}\\
&\quad\exp\left\lbrace -\frac{\zvec'\bm{S}_{\gamma}^{-1}\bm{S}_{\gamma}^{-1}\zvec}{2(1+g)}(1+g\,(1-
\tilde{R}_{\gamma,g}^2))\right\rbrace\exp\left\lbrace \frac{\zvec'\bm{S}_{\tilde\gamma}^{-1}\bm{S}_{\tilde\gamma}^{-1}\zvec}{2(1+g)}(1+g\,(1-
\tilde{R}_{\tilde\gamma,g}^2))\right\rbrace p(g)\mathrm{d}g,
\end{aligned}
\end{equation*}
where we call
\begin{equation*}
\begin{aligned}
\tilde{R}_{\gamma,g}^2 &=1-\frac{\zvec'\bm{S}_{\gamma}^{-1}(\mI-\bm{M}_{\gamma}\bm{M}_{\gamma}')\bm{S}_{\gamma}^{-1}\zvec}{\zvec'\bm{S}_{\gamma}^{-1}\bm{S}_{\gamma}^{-1}\zvec}
\end{aligned}
\end{equation*}
the implicit copula coefficient of determination, which (as opposed to the ordinary coefficient of determination)  depends on $g$ in the copula model.
\end{prop}
\vspace{-0.5em}
Proof of Proposition~\ref{prop:BF} can be found in Part~C of the Web Appendix, and
the expression for $\mbox{BF}(\gammavec | \widetilde\gammavec)$ involves
an integral which can be evaluated numerically. 
Part~C of the Web Appendix also
gives the Bayes factor in the special case where $\gammavec=\bm{0}$ (i.e. the empty model).

\section{Simulation Study}\label{sec:sim}
To 
illustrate the effectiveness of our approach
we undertake a simulation study. We employ the copula model at~(\ref{eq:copmod}) with non-parametric 
estimates of the margins $\hat F_Y$, and label this `BVSC' throughout the rest {of this Section. We consider the four
priors for $g$ outlined in Section~\ref{subsec:gen:idea} and compare the BVSC to two benchmark methods. The first is that 
of~\cite{LiaPauMolClyBer2008}, which has Gaussian disturbances and the same hyperpriors 
for $g$ (so that there are also four variants). To evaluate the posterior for this model we
use an MCMC sampler similar to that described in Section~\ref{subsec:posteval}, and label
this as `BVS'.
The second  benchmark is the approach of~\citet{RosRub2017} as implemented in the R-package \texttt{mombf} and labelled `mombf', where the product MOM (pMOM) non-local prior
proposed by~\cite{JohRos2012} 
is used.
 This provides estimates of the posterior model probabilities with normal (N), asymmetric normal (AN), Laplace (L) and asymmetric Laplace (AL) errors, and we label the benchmark by these distribution types. However, evaluation of the 
predictive distribution is only available for the normal error model.

\subsection{Simulation Design}\label{subsec:simdesign2}
We generate $n=200$ observations of  $p=20$ correlated covariates
$\xvec=(x_1,\ldots,x_{20})'\sim\ND(\nullvec,\mSigma)$, where $\mSigma=\bm{D}'\bm{D}$  and 
$\bm{D}$ is an upper triangular Cholesky factor with non-zero elements generated as N$(0,0.1^2)$.
The resulting $(n\times 20)$ design matrix $\bm{X}$ is then mean-centered (so that $\onevec'\bm{X}=\nullvec$), 
creating a correlated but numerically stable design. For $j=1,\ldots,20$, we set $\beta_j=0$ with probability 0.75, otherwise we generate $\beta_j$ from an
equally-weighted mixture of the two normals N$(1,0.25^2)$ and N$(-1,0.25^2)$. 
Setting $\betavec=(\beta_1,\ldots,\beta_{20})'$,  
observations of the dependent variable are generated from the following three distributions:
\begin{equation*}
\begin{aligned}
	&\mbox{Case 1, Normal: } &Y_{i} &= \xvec_{i}'\betavec+\varepsilon_{i},& \varepsilon_{i}&\sim\ND(0,r_1^2)\,,\\
&\mbox{Case 2, Log-normal: } &Y_{i} &= \exp\left(\xvec_{i}'\betavec+1.5\varepsilon_{i}\right), &\varepsilon_{i}&\sim\ND(0,r_2^2)\,,\\
	&\mbox{Case 3, Implicit Copula: } &Y_{i} &= F_{\LND}^{-1}(\Phi_1(z_i);-2.89,2),\, z_i=\xvec_{i}'\betavec+\varepsilon_{i},\, &\varepsilon_{i}&\sim\ND(0,r_3^2)\,,
\end{aligned}
\end{equation*}
for $i=1,\ldots,n$, and where $F_{\LND}$ is the log-normal distribution function.
The distribution in case~1 matches that 
of the Gaussian linear model (which is assumed in the BVS and mombf/N benchmarks), while that in case~3 matches that of the implicit copula model (i.e.~BVSC). 
The distribution in case~2 matches neither model.
For each of the three cases we simulated $K=100$ datasets using the same design matrix $\bm{X}$, which we refer to as replicates.
To make the three cases comparable, we set $r_1,r_2$ and $r_3$ to values that give a signal-to-noise
ratio (SNR) equal to 8; see the Part~E of the Web Appendix for details.

\subsection{Results}
To compare the approaches we consider two metrics. The first metric measures the accuracy of the predictive densities, and the second measures the correct selection of variables.  
\subsubsection{Prediction Accuracy}
To measure the accuracy of the predictive density of the dependent variable
we use the mean logarithmic score computed by ten-fold cross-validation.
For a given replicate, we compute this  by partitioning the data into ten equally sized sub-samples of sizes $n_k$,
denoted here as
$\{(y_{i,k},\xvec_{i,k});i=1,\ldots,n_k\}$ for $k=1,\ldots,10$.
For each observation 
in sub-sample $k$, we compute the predictive
density estimator at Eq.~(\ref{eq:phatyx}) using the remaining nine sub-samples as the training data, and denote these densities here
as $\hat p_k(y_{i,k}|\xvec_{i,k})$. The ten-fold mean logarithmic score is then
$\mbox{MLS}=\frac{1}{10}\sum_{k=1}^{10}\frac{1}{n_k}\sum_{i=1}^{n_k}\log \hat{p}_k(y_{i,k}|\xvec_{i,k})$.
\begin{sidewaysfigure}
	\begin{center}
		\includegraphics[width=0.96\textwidth,angle=0]{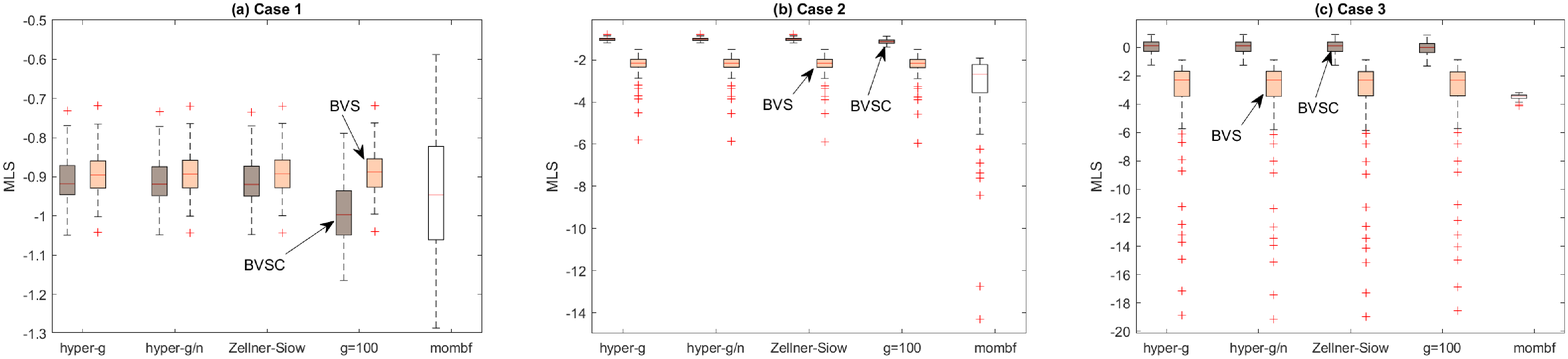}
	\end{center}
	\caption{Comparison of the predictive MLS from the simulation study. The three panels provide results for the three 
		cases. Each boxplot is of the 100 values of the MLS from the 100 simulation replicates, where higher values correspond
		to increased accuracy. In each panel, the first eight boxplots correspond to combinations of the methods
		BVSC and BVS with the four priors for $g$. The last boxplot (white) corresponds to the mombf/N method. BVSC dominates all other methods in the non-Gaussian
		cases~2 and~3. This figure appears in color in the
		electronic version of this article, and any mention of color refers to that version.}
	\label{fig:logcv}
\end{sidewaysfigure}

Figure~\ref{fig:logcv} gives boxplots of the MLS of the 100 replicates 
for the three cases in panels~(a--c). In each panel, nine methods are considered: the four 
variants of both BVSC and BVS, 
and mombf/N. 
We make three observations. First, the results for BVS are robust with respect to choice of prior for $g$,
whereas for BVSC fixing $g=100$ is dominated by the three flexible priors.
Second, for the Gaussian data in case~1 BVS is slightly better than BVSC,
and mombf/N has the highest variance. Third, BVSC outperforms both BVS and mombf/N substantially 
in the two non-Gaussian cases~2 and~3, which in case~3 is 
because the data is generated from a
copula model.

\subsubsection{Selection Accuracy}
The second measure is the precision-recall curve, which is a popular criterion for assessing
classification in machine learning. Here, the classification problem is the
selection from 20 covariates, 
using the marginal posteriors $\mbox{Pr}(\gamma_j=1|\yvec)$, $j=1,\ldots,20$, produced
by each method. Given a threshold probability value, let TP, FP and FN be true-positive, false-positive and false-negative classification rates, respectively. 
Then the curve plots Recall=TP/(TP+FN) on the horizontal
axis, against Precision=TP/(TP+FP) on the vertical axis, as the threshold probability value varies from 0 to 1. 
Simultaneously high values of Recall and Precision (i.e. curves that look like a transposed letter `L') indicate
accurate classification.
\begin{sidewaysfigure}[p]
	\includegraphics[width=0.96\textwidth,angle=0]{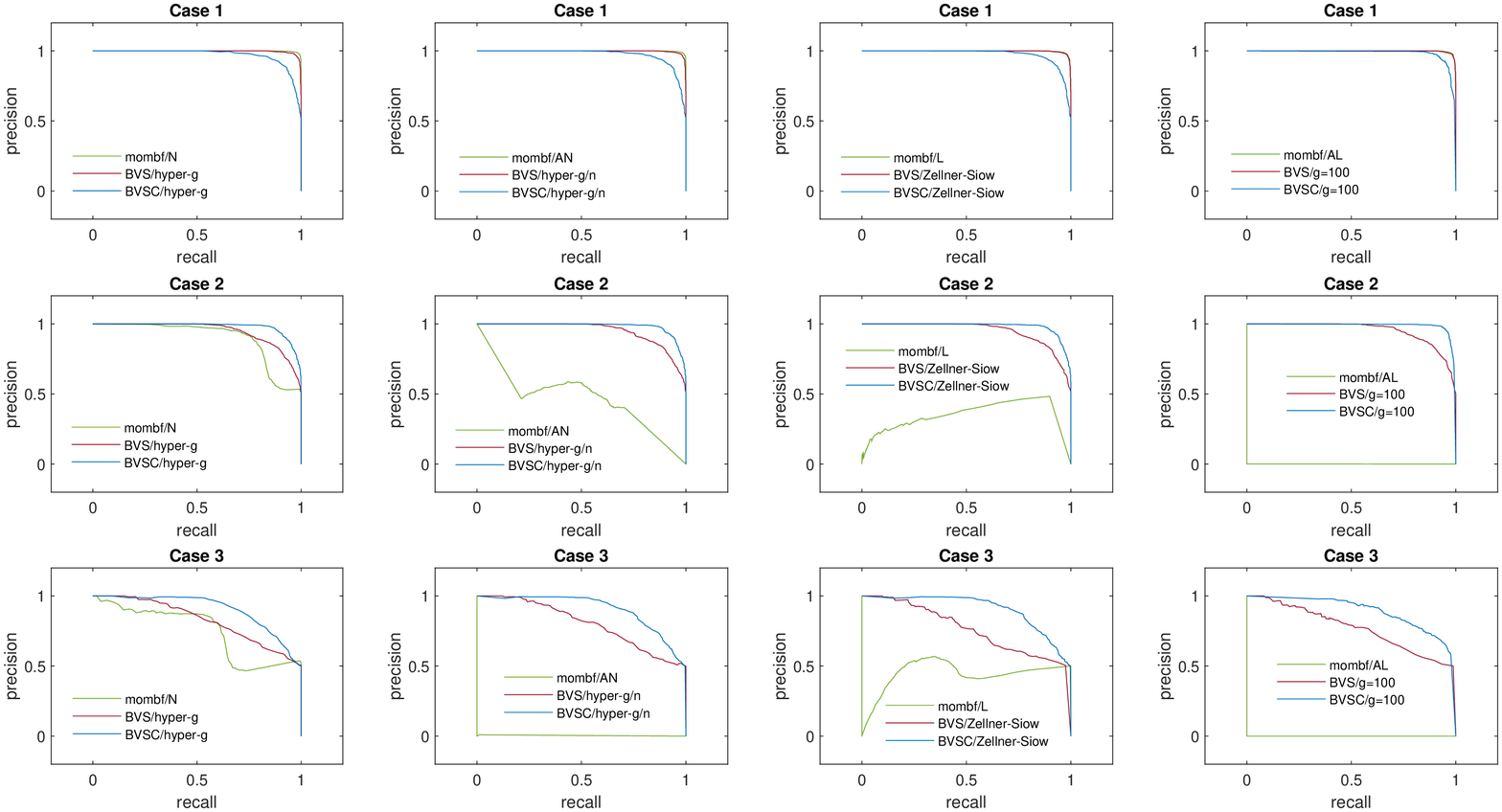}
	\caption{Comparison of the average recall-precision curves from the simulation study for a grid of thresholds in $(0,1)$.
		The first row (i.e. panels~a,b,c,d) show results for case~1, the second row (i.e. panels e,f,g,h) for case~2, and the third row (i.e. panels i,j,k,l) for case~3. Each panel contains curves for the BVSC (blue) and the BVS (red) methods for one specific prior for $g$, along with curves from mombf (green) with one distributional assumption from N (normal), AN (asymmetric normal), L (Laplace) and AL (asymmetric Laplace) in each column. Curves with simultaneously higher 
		recall and precision are more accurate classifiers. This figure appears in color in the
		electronic version of this article, and any mention of color refers to that version.}
	\label{fig:prec:recall}
\end{sidewaysfigure}

Figure~\ref{fig:prec:recall} plots the average precision-recall curves over the 100 replicates
of the simulation.
For each case, 12 curves---one for each method and variant---are presented, and
we make three observations.
First,  the asymmetric variants of mombf perform
 poorly for the non-Gaussian data (cases 2 and 3),
despite specifically allowing for this circumstance. 
Second, the BVSC is much more accurate in cases 2 and 3, where it also outperforms
BVS for all variants of priors for $g$. This robustness to distributional form 
is a result of the flexibility obtained through the non-parametric calibration of the margins.
Third, BVSC is less accurate than  
BVS and mombf only for the Gaussian data generating process (case 1), although the under-performance is minor when compared to the 
gains obtained in the two non-Gaussian cases.
\section{Extension to Spatial BVS for fMRI}\label{sec:fmri}
An important application of BVS is to construct activation maps in 
functional magnetic resonance imaging (fMRI) studies. This involves
the extension to spatial data located on a
regular lattice of `voxels' that partition the brain~\citep{SmiPueAueFah2003,SmiFah2007,goldsmith2014,lee2014}. 
We show how to extend our methodology to this case, and demonstrate that 
the activation maps
can be more accurate when taking into account the non-Gaussianity of the magnetic resonance (MR) signal using our implicit copula-based approach to variable selection. This is consistent with 
recent work that suggests the MR signal is neither conditionally
Gaussian nor homoscedastic~\citep{eklund2017}.
\subsection{Marginally calibrated variable selection for fMRI}\label{subsec:fmri:cali}
In fMRI studies, a MR signal  
$\{Y_{i,t}; t=1,\ldots, T\}$ is observed at each voxel $i\in\lbrace 1,\ldots,N\rbrace$.
This is matched with a series of scalar observations $\{x_{i,t}; t=1,\ldots,T\}$ on a transformed
stimulus,
which is a delayed and continuously modified version of an original stimulus 
called the `hemodynamic response' that is derived in a pre-processing step outlined in~\citet[p.804]{SmiPueAueFah2003}. The objective in such analyses is to identify at which
voxels the MR signal is related to the transformed stimulus~\citep{BezHugGal2018}. 

Denote voxel activation by the vector $\gammavec=(\gamma_1,\ldots,\gamma_N)'$, such that voxel $i$ 
is activated by the stimulus if $\gamma_i=1$, and inactivate if $\gamma_i=0$. 
Spatial
smoothing is essential to obtaining 
reliable activation maps, which is achieved by using the mass function of an Ising model as a prior
\[
p(\gammavec|\theta)\propto\exp\left(\sum_{i=1}^N\delta_i\gamma_i+\theta\sum_{i\sim j}\omega_{ij}{\cal I}(\gamma_i=\gamma_j)\right)\,.
\]
Here, ${\cal I}(A)=1$ if $A$ is true and zero otherwise, and the
summation over $i\sim j$ is over all pairwise neighboring elements
of $\gammavec$ (which
are the up to 8 neighbors of each voxel in two dimensions). 
The weight $\omega_{ij}=1$ for immediately adjacent voxels $i,j$, and $\omega_{ij}=1/\sqrt{2}$ for 
diagonally adjacent voxels $i,j$, while
the parameter $0\leq \theta \leq 0.45$ controls the level of
spatial smoothing. 
The $\delta_1,\ldots,\delta_N$ are coefficients of the external field and
determined using a process described in~\citet[Sec.4.2]{SmiFah2007} that
computes each $\delta_i$ from the amount of grey matter in voxel $i$ (with more
grey matter resulting in a higher value of $\delta_i$). 
This is important
because only grey matter can be activated by the stimulus.
Alternative binary random fields may also be used as a prior
for $\gammavec$ in our framework.

To extend our copula model in Section~\ref{subsec:margcal} to this case, let $\bm{Y}_i=(Y_{i,1},\ldots,Y_{i,T})'$, $\xvec_i=(x_{i,1},\ldots,x_{i,T})'$, $\mY=(\mY_1',\ldots,\mY_N')'$ and $\xvec=(\xvec_1',\ldots,\xvec_N')'$. Also, let 
$\wvec_t=(w_{t,1},\ldots,w_{t,m})'$ be $m$ basis functions (which we specify later) evaluated at time point $t$ used to capture a localized baseline time 
trend at each voxel, and 
$\bm{W}=\lbrack \wvec_1|\ldots|\wvec_T\rbrack'$.
We assume $Y_{i,t}$ has a voxel-specific marginal distribution
function $F_{Y_i}(y_{i,t})$ and density $p_{Y_i}(y_{i,t})$, and
adopt the following copula model for the joint density of 
$\mY|\xvec,\bm{W},\gammavec$: 
\begin{equation}\label{eq:copmod:fmri}
p(\yvec|\xvec,\bm{W},\gammavec)=\prod_{i=1}^N \left(c_{\mbox{\tiny SBVS}}(\uvec_i|\xvec_i,\bm{W},\gamma_i)\prod_{t=1}^T p_{Y_i}(y_{i,t})\right)\,,
\end{equation}
where $\uvec_i=(u_{i,1},\ldots,u_{i,T})'$ and $u_{i,t}=F_{Y_i}(y_{i,t})$.
At~\eqref{eq:copmod:fmri}, the MR signals $\mY_1,\ldots,\mY_N$ are independent over 
voxels when conditioning on $\gammavec$, with each vector $\bm{Y}_i$ following
a copula decomposition as at~\eqref{eq:copmod}. For
the $T$-dimensional copula density $c_{\mbox{\tiny SBVS}}$ we employ an implicit copula
derived from a pseudo-regression model, as outlined below in Section~\ref{subsec:SBVSC}. 

Spatial
dependence between elements of $\gammavec$ is introduced using the Ising
model prior, giving posterior mass function
$p(\gammavec|\xvec,\bm{W},\yvec)\propto \left(\prod_{i=1}^N c_{\mbox{\tiny SBVS}}(\uvec_i|\xvec_i,\bm{W},\gamma_i)\right) p(\gammavec|\theta)$.
As in Section~\ref{sec:implicit:copula}, variable selection (i.e. the classification of voxels as active or inactive) using
this posterior
is separated from the task of marginal calibration of the distributions
$F_{Y_1},\ldots,F_{Y_N}$ of the  MR signal. 
We show in our empirical work 
that this changes the activation maps and improves the quality of fit (measured
using mean logarithmic scores) substantially compared to the Gaussian spatial BVS of~\cite{SmiPueAueFah2003} and \cite{SmiFah2007}.
\subsection{Spatial variable selection copula}\label{subsec:SBVSC}
The implicit copula $c_{\mbox{\tiny SBVS}}$ is of the same form at every voxel,
and we derive it here for voxel $i$. 
It is obtained from the regression $\tilde Z_{i,t}=\wvec_{t}'\alphavec +  x_{i,t}\beta_{\gamma_i}+\varepsilon_{i,t}$ for pseudo-response $\tilde Z_{i,t}$
at times $t=1,\ldots,T$. Here,
 $\varepsilon_{i,t}\overset{}{\sim}\ND(0,\sigma^2)$ and $\wvec_t'\alphavec$ is a time trend with eight low order Fourier terms as basis functions and coefficients $\alphavec$, while the transformed stimulus $x_{i,t}$ is a scalar covariate
 with coefficient $\beta_{\gamma_i}$. 
Variable selection is performed only on $x_{i,t}$, so that 
$\beta_{\gamma_i}=0 \mbox{ iff } \gamma_i=0$. 
\cite{SmiPueAueFah2003} and~\cite{SmiFah2007} apply regressions of this form
directly to the MR signal at each voxel, but here we instead extract its implicit copula for use at~\eqref{eq:copmod:fmri}.

Setting $\widetilde\mZ_i=(\tilde Z_{i,1},\ldots,\tilde Z_{i,T})'$, the regression can be written as the linear model
\begin{equation}\label{eq:frmi}
\widetilde\mZ_i= \bm{W}\alphavec + \xvec_i\beta_{\gamma_i}+\varepsilonvec_i\,,
\end{equation}
with $\varepsilonvec_i=(\varepsilon_{i,1},\ldots,\varepsilon_{i,T})'$, and
$\xvec_i, \bm{W}$ as defined
in the previous subsection. 
Proper priors have to be employed for $\alphavec$ and $\beta_{\gamma_i}$ to obtain a proper implicit copula.
We use the same spike-and-slab prior for $\beta_{\gamma_i}$
as previously, so that
$\beta_{\gamma_i}=0|\gamma_i=0$, and 
$\beta_{\gamma_i}|\gamma_i=1,g,\sigma^2 \sim \ND\left(0,g\sigma^2(\xvec_i'\xvec_i)^{-1}\right)$.
We use a $\ND(0,\sigma^2 d \mI)$ prior for $\alphavec$, with $d$ set to make the prior uninformative, along with the same four priors for $g$ used previously in Section~\ref{sec:implicit:copula}.

The same process is used to construct $c_{\mbox{\tiny SBVS}}$ as in Section~\ref{subsec:gen:idea}, but tailored to account for the time trend in~\eqref{eq:frmi}.
First, the joint distribution 
$\widetilde\mZ_i| \xvec_i,\bm{W},\gamma_i,g,\sigma^2\sim N(\bm{0},\mOmega)$, with
\[\mOmega =  \sigma^2\left( \mI + d \bm{W} \bm{W}' + \gamma_i\left(\frac{g  \xvec_i \xvec_i'}{\xvec_i'\xvec_i}\right)\right)\,,
\]
where $\beta_{\gamma_i}$ and $\alphavec$ have been integrated out analytically; see Part~G.1 of the Web Appendix. 
Second, the pseudo-response is standardized by the marginal variances of this distribution 
to give $\bm{Z}_i=\sigma^{-1}\bm{S}_{\gamma_i}\widetilde{\bm{Z}}_i$, with
$\bm{S}_{\gamma_i}\equiv\diag(s_{1}(\gamma_i),\ldots,s_{T}(\gamma_i))$ and 
$
s_t(\gamma_i)=\left(1 + d \wvec_{t}'\wvec_{t} +  \gamma_i\left(\frac{g x_{i,t}^2}{\xvec_i'\xvec_i}\right)\right)^{-\frac{1}{2}}
$.
Thus, 
$\mZ_i|\xvec_i,\bm{W},\gamma_i,g,\sigma^2\sim \ND\left(\bm{0},\bm{R}( \xvec_i,\bm{W},\gamma_i,g)\right)$, where \begin{equation}
\label{eq:correlationmatrix2}
\bm{R}(\xvec_i,\bm{W},\gamma_i,g) = \bm{S}_{\gamma_i}\left( \mI + d \bm{W} \bm{W}' + \gamma_i \frac{g \xvec_i\xvec_i'}{\xvec_i'\xvec_i}\right) \bm{S}_{\gamma_i}\,,
\end{equation}
is a correlation matrix, and the implicit copula of 
both $\widetilde{\bm{Z}}_i$ and $\bm{Z}_i$ 
(conditional on $\xvec_i,\bm{W},\gamma_i,g,\sigma^2$) 
is a Gaussian copula with parameter matrix $\bm{R}$ defined at~\eqref{eq:correlationmatrix2}.
Finally, we obtain $c_{\mbox{\tiny SBVS}}$ by mixing over the prior
for $g$, which we formalize in the following definition.

\begin{defin}
\label{defin:scbvs}
Let $c_{\mbox{\tiny{Ga}}}$ be 
the Gaussian copula density with the correlation matrix $\bm{R}$ defined at~\eqref{eq:correlationmatrix2}. 
If
$p(g)$ is a proper prior density for $g>0$, then we call
\[
c_{\mbox{\tiny SBVS}}(\bm{u}|\xvec_i,\bm{W},\gamma_i)=\int c_{\mbox{\tiny{Ga}}}(\bm{u};\bm{R}(\xvec_i,\bm{W},\gamma_i,g))p(g)\mathrm{d}g\,.
\]
the density function of a spatial variable selection copula.
\end{defin}
\noindent 
As in Section~\ref{sec:implicit:copula}, $\sigma^2$
does not feature in the expression for $\bm{R}$ nor the copula density. We use this copula at~\eqref{eq:copmod:fmri}, where the 
transformed stimulus values $\xvec_i$ and activation indicator $\gamma_i$ vary over voxels, whereas
the matrix of linear time trend basis terms $\bm{W}$ does not. 
	
\subsection{Estimation and inference}
Parameter estimation and posterior inference is obtained using 
an adaptation of the sampler in Section~\ref{subsec:posteval} that 
is presented in detail in Parts~G.2 and G.3 of the Web Appendix. It is a single-site sampler in the elements of $\gammavec=(\gamma_1,\ldots,\gamma_N)'$, 
which is more computationally
efficient than multi-site samplers (e.g.~\cite{nottgreen2004}) because 
key quantities can be pre-computed just once. 
This is important
when undertaking variable selection 
in fMRI studies due to the large number of voxels $N$.
The parameter $g$ is generated
using a Metropolis-Hastings (MH) step with a Gaussian approximation as a proposal.
This has an acceptance rate of over 80\%, and replaces
the HMC step in Section~\ref{sec:estimation}. The
spatial smoothing parameter $\theta$ is generated using an adaptive random
walk MH step. 

Accurate Monte Carlo mixture estimates of the marginal posteriors $\mbox{Pr}(\gamma_i=1|\xvec,\bm{W},\yvec)$ can be readily computed from 
the conditional posteriors of the single site sampler. Plots of these are Bayesian estimates of the 
activation maps. An accompanying output is the map of posterior means of the 
amplitudes $\beta_{\gamma_1},\ldots,\beta_{\gamma_N}$, for which an efficient mixture estimate can 
also be constructed, as outlined in Part~G.4 of the Web Appendix.
\begin{table}
	\caption{Results from applying both the copula (SBVSC) and Gaussian (SBVS) spatial Bayesian variable selection methods to the fMRI data.
		The four different priors are used for $g$, giving a total of eight methods. The top rows report the mean logarithmic scores (MLS) of the MR signal (multiplied by 100 for presentation), broken down voxels classified as active,
		inactive and overall. Higher MLS values indicate greater accuracy. The bottom rows report the posterior mean and standard deviation of the number of active voxels
		$q_\gamma=\sum_{j=1}^N\gamma_j$.}\label{tab:logsc:fmri}
	\begin{tabular}{lcccccccc}
		\hline
		\hline
		Prior  & \multicolumn{2}{c}{ hyper-g} & \multicolumn{2}{c}{ hyper-g/$n$} & \multicolumn{2}{c}{ Zellner-Siow} & \multicolumn{2}{c}{ $g=100$}\\
		\hline
		Model  & SBVSC &  SBVS    & SBVSC &  SBVS   & SBVSC &  SBVS   & SBVSC &  SBVS\\
		\hline
		Active  & -411.65 &-450.72  & -411.65 & -450.72 & -411.65 & -450.72 & -418.37 &  -451.08\\
		Inactive & -408.12 & -416.48 & -408.12 &  -416.48  &-408.12 &  -416.48  &-408.32 & -416.53\\
		Overall &  -408.18 & -416.92 & -408.18 & -416.92 & -408.17 & -416.92 &  -408.42 &-416.96\\\hline
		$\dsE(q_\gamma|\yvec)$ & 110.8 & 162.0 & 110.8 & 162.0 & 110.8 & 162.0 & 72.1 & 119.5\\
		$\mbox{Std}(q_\gamma|\yvec)$ & 3.1 & 13.9 & 3.1 & 13.9 & 3.1 & 13.9 & 2.3 & 8.6\\
		\hline\hline
	\end{tabular}
\end{table}
\begin{figure}
	\centering\includegraphics[width=0.9\textwidth,angle=0]{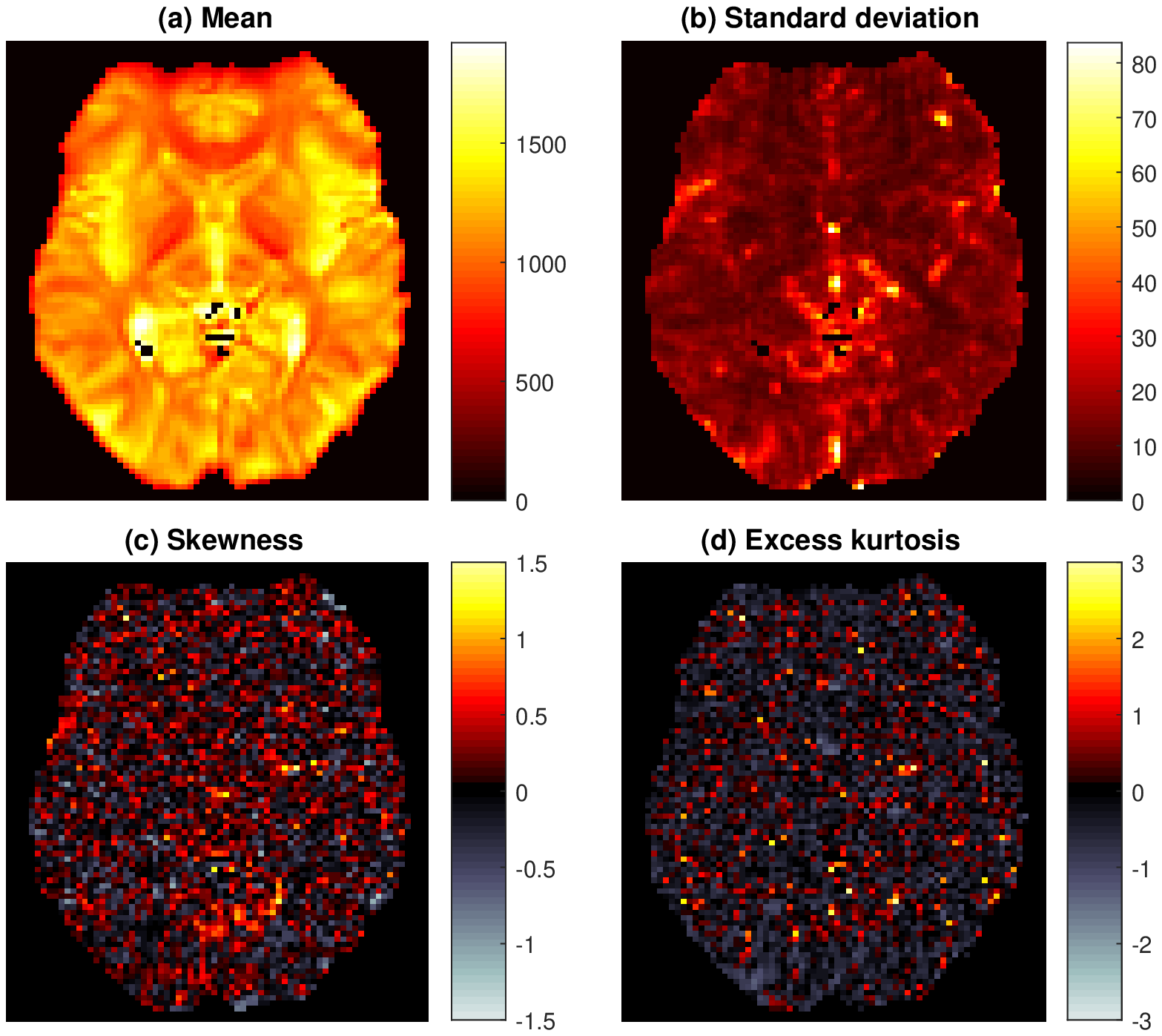}
	\caption{Sample moments of the MR signal at each voxel in the fMRI example. The four panels plot the (a) mean, (b) standard deviation, (c) Pearson skew and (d) excess kurtosis values for each voxel. Computation of the global Moran's I spatial correlation coefficients indicates that there is strong spatial correlation in all four sample
		moments. This figure appears in color in the
		electronic version of this article, and any mention of color refers to that version.}
	\label{fig:fmri:summaries}
\end{figure}

\subsection{Empirical results} 
To illustrate, we construct activation maps for slice 10 of individual B in~\citet{SmiPueAueFah2003}. This data includes an MR signal observed at 
$T=63$ time points and $N=72\times 86$ voxels from a simple visual experiment. Activation is therefore 
largely in the 
visual cortex. 
Figure~\ref{fig:fmri:summaries} plots the first four sample moments of the
MR time series at each voxel, indicating a considerable deviation from normality
at many voxels. This is not captured in the Gaussian 
spatial BVS model (labelled `SBVS' here). 
In contrast, our spatial BVS {\em copula} model (labelled `SBVSC' here)
does so using non-parametric estimators for $F_{Y_1},\ldots,
F_{Y_N}$.

We estimate the SBVSC parameters for all four priors for $g$.
For comparison, we also estimate the SBVS model of~\cite{SmiPueAueFah2003} using 
the same priors for $g,\theta,\gammavec$ as in the copula model. To compare
the two sets of results,  
Table~\ref{tab:logsc:fmri} reports the mean (in-sample) logarithmic 
scores for 
both the SBVS and SBVSC models, and for each prior of $g$. To compute these
we evaluate the mean in-sample logarithmic scores at each voxel $i$ as 
$1/T\sum_{t=1}^T \log(\hat p(y_{i,t}|\xvec))$, and
then average the results across active, inactive and all voxels. The predictive
densities are computed using a minor adjustment to~\eqref{eq:phatyx} to account
for the time trend terms; see Part~G.5 of the Web Appendix. We
make three observations. First, in all cases the SBVSC scores are 
higher than those for SBVS. The relative improvement is 9.5\% for voxels classified
as active by SBVSC, and 2\% across all voxels.
Second, for SBVSC setting $g=100$ degrades the results, although there
is no difference between the other three priors for $g$. 
Third, the expected number of active voxels is lower for SBVSC, producing
a sparser image than SBVS.
\begin{sidewaysfigure}
	\centering\includegraphics[width=0.7\textwidth,angle=0]{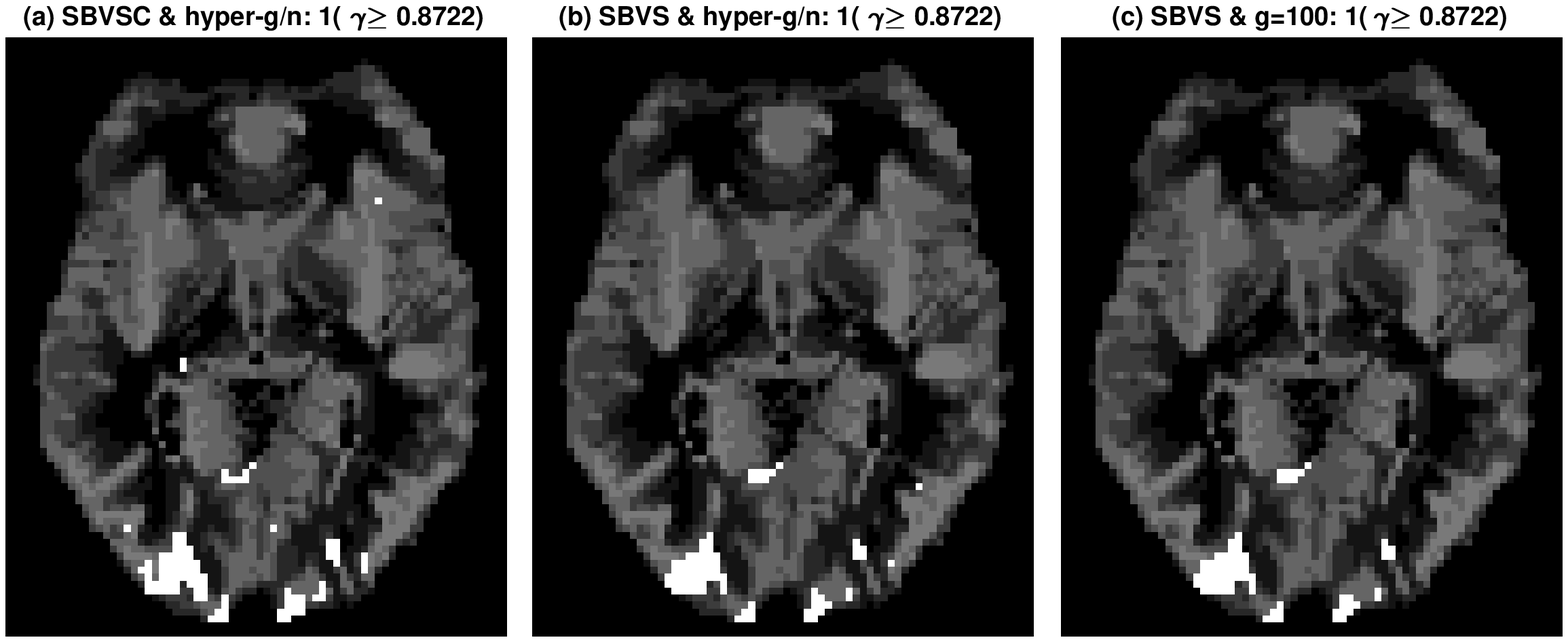}\\\vspace{0.1cm}
	\includegraphics[width=0.7\textwidth,angle=0]{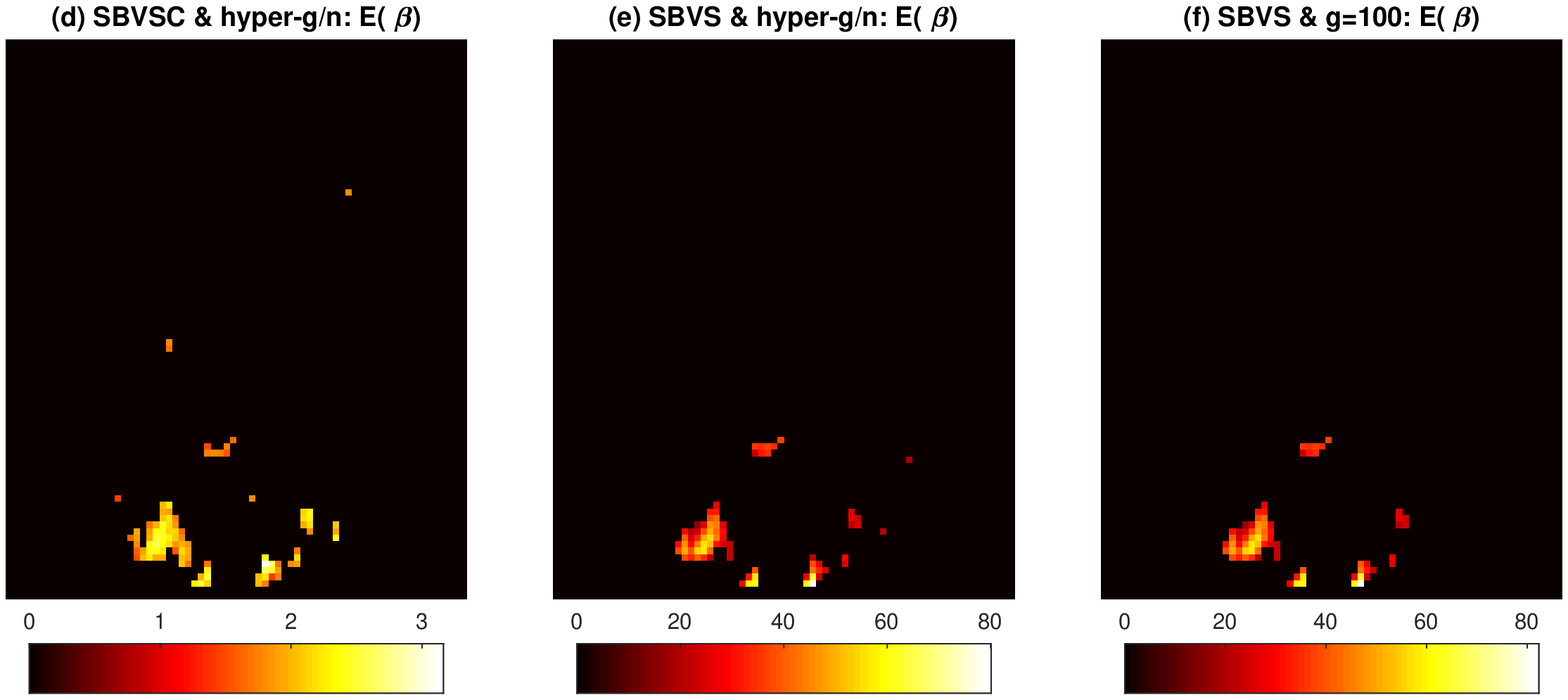}
	\caption{Posterior activation and amplitude maps for the fMRI data from three
		different estimators. The first row shows the activation maps (where white
		voxels are those classified as active), and the second shows
		the mean activation amplitudes. The three different estimators are: (a,d) SBVSC with hyper-g/n prior; (b,e) SBVS with hyper-g/n prior; and, (c,f) SBVS with $g=100$. This figure appears in color in the electronic version of this article, and any mention of color refers to that version.}
	\label{fig:fmri:gamma}
\end{sidewaysfigure}
\begin{figure}[t]
	\centering\includegraphics[width=0.5\textwidth,angle=0]{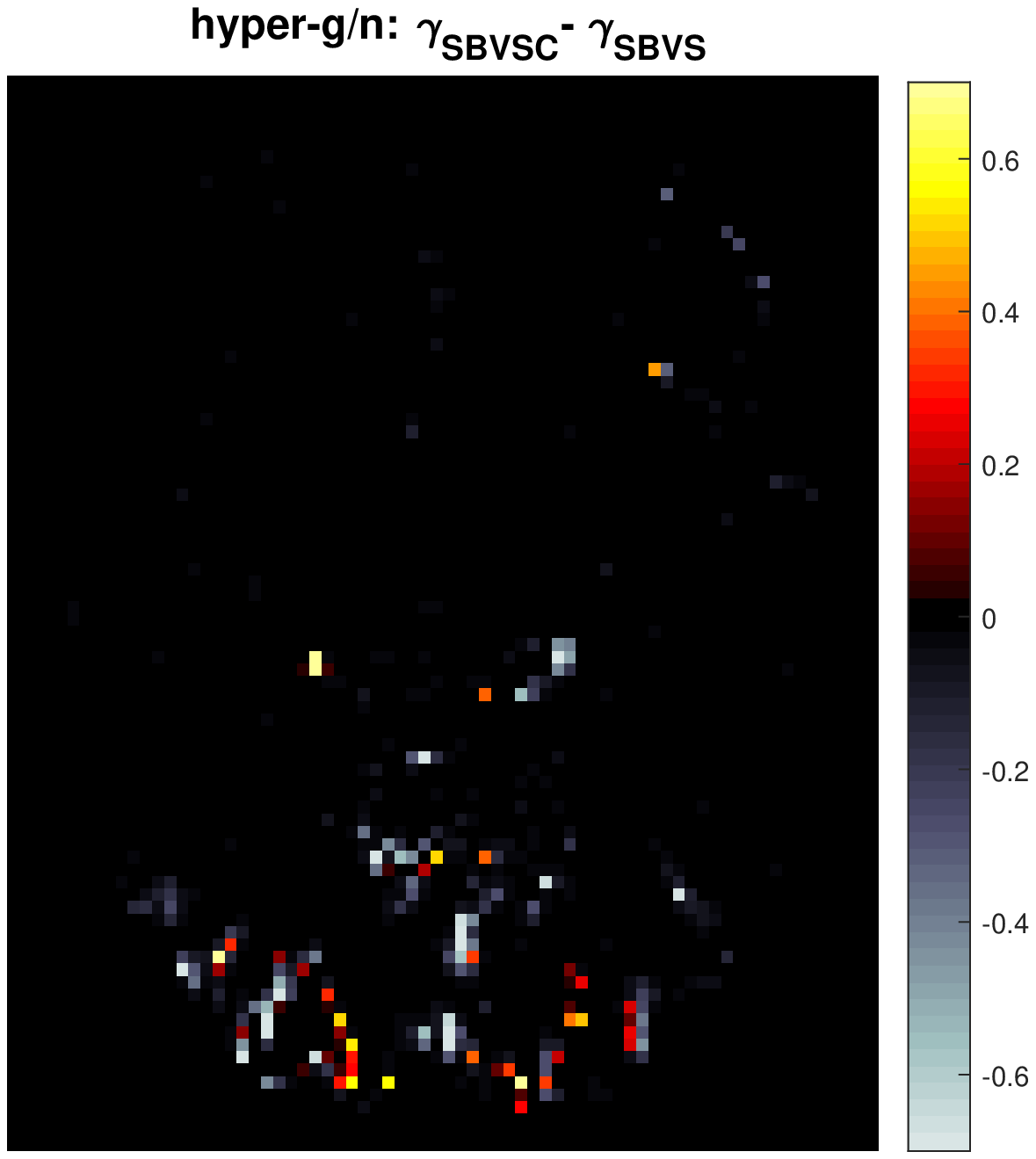}
	\caption{Difference between the activation probabilities of the 
		copula (SBVSC) and Gaussian (SBVS) spatial Bayesian variable selection models for 
		the fMRI data. The difference is SBVSC minus SBVS, and both 
		models employ the same Ising prior for $\gammavec$ and hyper-g/n prior for $g$. This figure appears in color in the
		electronic version of this article, and any mention of color refers to that version.}
	\label{fig:fmri:diffgamma}
\end{figure}

Based on these observations, Figure~\ref{fig:fmri:gamma} compares
activation
and amplitude maps for three cases: panels~(a,d) SBVSC with hyper-g/n prior;
(b,e)~SBVS with hyper-g/n prior; and~(c,f) SBVS with $g=100$. The latter 
case is included because it is the benchmark model suggested by~\cite{SmiFah2007}. The activation maps in the first row are 
obtained by defining a voxel as active if and only if
$\mbox{Pr}(\gamma_i=1|\xvec,\bm{W},\yvec)\geq 0.8722$. The justification 
	for this threshold value is that $-2\log((1-\mbox{Pr}(\gamma_i=1|\xvec,\bm{W},\yvec))/\mbox{Pr}(\gamma_i=1|\xvec,\bm{W},\yvec)$ is approximately $\chi^2(1)$ distributed and the threshold corresponds to a p-value of 0.05. 
The maps for the two SBVS models are close to identical,
but differ from those for the SBVSC model, with the latter allowing for 
sharper edges in the amplitude maps. To further 
highlight this difference, Figure~\ref{fig:fmri:diffgamma}
plots the difference in the activation probabilities between the copula
and Gaussian models. These differ between -0.6 and 0.6, so that 
allowing for
more accurate marginal calibration of the MR signal not only increases
the logarithmic scores, but affects activation maps, which are the primary
output of fMRI processing.

Both samplers were implemented efficiently in MATLAB. The time to
undertake 1000 sweeps was approximately 10 mins (SBVS) and 11 mins 
(SBVSC) when $g$ is generated, and 2 mins (SBVS) and 2.3 mins (SBVSC) when $g$ is fixed. 
Thus, marginally-calibrated spatial BVS is only slightly slower than Gaussian spatial BVS,
and can be made even faster using a lower level language.
Typically, only a few thousand sweeps are needed to obtain highly accurate maps, because the 
Markov chain converges quickly and mixture estimators are used. While not undertaken here, the speed of the approach allows application to multiple slices using the 3D Markov random field prior for $\gammavec$ in~\cite{SmiFah2007}.
\section{Discussion}\label{sec:discussion}
This paper proposes a new tractable and general 
approach to undertake BVS for non-Gaussian
data. It uses a copula decomposition that
allows the marginal distribution of the dependent variable to be
calibrated accurately using nonparametric estimation. However,
we note that this does not imply the
other forms of calibration or dispersion listed by~\cite{gneiting2013}.
In addition, a referee pointed out that the assumption at~\eqref{eq:copmod} that $F_Y$ is independent of $\bm{X}$
limits the range of distributions for $\bm{Y}|\bm{X}$ that can be represented.
However, this assumption separates the variable selection problem from
that of marginal calibration, which is a major advantage of our proposed approach.

The key ingredient of our methodology 
is a family of implicit copulas that are constructed from
a regression model for a Gaussian pseudo-response with spike-and-slab priors. 
These mix over the different priors in~\cite{LiaPauMolClyBer2008} for the scaling factor 
of the g-prior, producing the copula family at Defn.~\ref{defin:cbvs} which is a 
continuous mixture of Gaussian copulas.
We apply our approach to an example with 6192 spatially correlated indicators, and to second example
with $p=252$ correlated covariates in Part~F of the Web Appendix. 
The empirical work demonstrates that mixing over the priors for $g$ (particularly the 
hyper-g prior) results in much
more accurate covariate selection and predictive densities of the dependent variable,
compared to fixing $g$. This is in contrast to Gaussian Bayesian variable selection, where in some examples fixed $g$ (such as $g=100$ or $g=n$) can provide good
results~\citep{SmiKoh1996}. Moreover, in the second example we also find that integrating out 
uncertainty in $F_Y$ as in~\cite{GraLis2017} does not meaningfully effect covariate selection or prediction.
The method is scalable to higher dimensions because estimation by
stochastic search over $\gammavec$ is fast when exploiting the
matrix identities in the Web Appendix. The extension of the method 
to spatial BVS for fMRI---an important contemporary 
application~\citep{lee2014,BezHugGal2018}---highlights its tractability.

While the use of copulas to capture dependence between multiple observations on 
one or more variables is rare, there are some recent examples. 
In regression these include implicit copulas constructed
from Gaussian processes~\citep{Wilson2010,Wauthier2010} or regularized basis functions~\citep{KleSmi2019}. In time series analysis,
copulas have been used to capture serial dependence in univariate
and multivariate data; for example, see~\cite{Smi2015} and \cite{shi2018}.
These studies all exploit a copula decomposition to allow
for non-Gaussian margins. However, ours is the first study to employ
our proposed copula formulation to undertake variable selection.

We finish by making some suggestions for future work. 
A potential extension is to multivariate regression (i.e. with multiple
dependent variables), generalizing the approaches of~\cite{brown1998}, 
\cite{SmiKoh2000} and others.
To do so requires the specification of
an implicit copula analogous to that in Defn.~\ref{defin:cbvs}.
It would also be interesting to explore the relationship between properties
of $c_{\mbox{\tiny BVS}}$, such as the dependence metrics in Part~C of the Web Appendix, and covariate selection accuracy.
Last, 
the
implicit copula can be extended to other Bayesian models for fMRI data that are based on alternative 
specifications of~\eqref{eq:frmi}, such as that in~\cite{lee2014}. This would 
lead to other copula processes on the covariate space with strong potential to further improve 
voxel classification accuracy.

\section*{Acknowledgements}
Nadja Klein gratefully acknowledges funding from the Alexander von Humboldt foundation and financial support through the Emmy Noether grant KL 3037/1-1 of the German research
foundation (DFG).
The authors thank two referees and an associate editor for extensive comments that
improved exposition in the manuscript.

\section*{Data Availability} 
The data that support the findings in this paper are available in the Supporting Information.
\setlength{\bibsep}{0pt plus 0.3ex}
\bibliography{litliste}
\section*{Supporting Information}
Online Appendices, Tables, and Figures referenced in Sections~1 through 5 are available with this paper at the Biometrics website on Wiley Online Library. MATLAB code to 
reproduce the results for the fMRI data in Section~5 is also available at the same website.
\end{document}